# CEP-stable soliton-based pulse compression to 4.4 fs and UV generation at 800 kHz repetition rate

ALEXEY ERMOLOV,[1,*] CHRISTIAN HEIDE,[2] PHILIP DIENSTBIER,[2] FELIX KÖTTIG,[1] FRANCESCO TANI,[1] PETER HOMMELHOFF,[1,2] AND PHILIP ST.J. RUSSELL[1,2]

[1]Max Planck Institute for the Science of Light, Staudtstr. 2, 91058 Erlangen, Germany
[2]Department of Physics, Friedrich Alexander University, Staudtstr. 1, 91058 Erlangen, Germany
*Corresponding author: alexey.ermolov@mpl.mpg.de

**We report generation of a femtosecond supercontinuum extending from the ultraviolet to the near-infrared and detection of its carrier-envelope phase variation by *f*-to-2*f* interferometry. The spectrum is generated in a gas-filled hollow-core photonic crystal fiber where soliton dynamics allows CEP-stable self-compression of OPCPA pump pulses at 800 nm to a duration of 1.7 optical cycles, followed by dispersive wave emission. The source provides up to 1 µJ of pulse energy at 800 kHz repetition rate resulting in 0.8 W of average power, and can be extremely useful for example in strong-field physics, pump-probe measurements and ultraviolet frequency comb metrology.**

Phase-stable laser pulses with energies in the µJ range and durations of a few fs or less are a powerful tool for studying and controlling electron dynamics and charge migration in matter and biological systems. In strong field physics, the carrier envelope phase (CEP) becomes increasingly important as the pulse duration approaches the single cycle limit. A stable CEP is, for example, crucial for the generation of isolated attosecond pulses [1,2], and can be exploited to break the inversion symmetry of graphene and generate CEP-dependent currents [3]. High repetition rates are of great importance for data acquisition at high signal-to-noise ratios and are especially important for studies of processes with low interaction cross-sections.

In the near and mid-infrared spectral regions, CEP stable laser pulses with few-cycle durations and µJ-level energies can be produced at hundreds of kHz repetition rates using an optical parametric chirped pulse amplifier (OPCPA). Additional pulse compression is however needed to approach the single cycle limit [4].

Realizing a system with a similar specification in the ultraviolet (UV) spectral range is considerably more challenging but also of great interest both for ultrafast spectroscopy and frequency metrology [5–7].

Here we report the generation of 1.7 optical cycles pulses with stable CEP at 800 kHz repetition rate, obtained via soliton self-compression in a gas-filled kagomé-type hollow-core photonic crystal fiber (kagomé-PCF). Such pulses are extremely useful for controlling electrons at optical frequencies inside solids and at surfaces, as demonstrated by electron emission from a tungsten nanotip [8] as well as by coherent electron trajectory control in graphene [9].

The same system can be used to generate wavelength-tunable UV radiation at high average powers via dispersive wave (DW) emission [10], with pulse durations of a few fs [11,12] when appropriate dispersion compensation is applied. Even though the CEP stability of DW emission has been demonstrated in silicon waveguides [13], it has never been characterized in a gas-based system, where the strong shock effect and ionization may play a detrimental role. Here, we report measurements of the CEP variation of the DW pulses and show that they exhibit a phase stability similar to the self-compressed higher order solitons from which they are emitted. As a result, the source is suitable for pump-probe phase sensitive UV spectroscopy and may open a new route for realizing frequency combs in the deep UV.

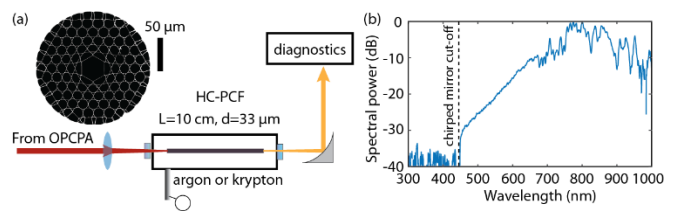

Fig. 1. (a) Schematic of experimental setup. The elements are described in the main text. Inset: scanning electron micrograph of the kagomé-PCF used in the experiment. (b) Experimental spectrum after propagation of 1 µJ pump pulses through 10 cm of the PCF filled with 3.9 bar of argon. The cut-off at 450 nm is due to the bandwidth limit of the chirped mirrors placed after the output of the PCF.

A sketch of the experimental apparatus is shown in Fig. 1(a). The pump laser is an OPCPA from Venteon seeded by an actively stabilized Ti:Sa oscillator [14]. The repetition rate can be switched between 200, 400 and 800 kHz. The feedback from an *f*-2*f* interferometer at the amplifier output can be used to stabilize the slow drift of the carrier envelope offset (CEO). Because of limited pulse energy, however, the system could only be stabilized at 200 kHz repetition rate. Operating at 800 kHz, the system delivers 9 fs pulses with 3.5 µJ pulse energy (2.8 W average power). In the experiments we used a 10-cm-long kagomé-PCF with 33 μm hollow core diameter (a scanning electron micrograph is shown in the inset of Fig. 1(a)), placed inside a gas cell equipped with 1.5 mm thick uncoated $MgF_2$ windows at both fiber ends.

The fiber was pumped by pulses from the OPCPA, and the output light was collimated using an aluminum-coated 90° off-axis parabolic mirror. While the pulses are self-compressed at the fiber output, it was necessary to compensate for the dispersion introduced by the gas cell output window and ~2 m of propagation in air using ultra-broadband chirped mirrors (–40 $fs^2$ per bounce, Ultrafast Innovation PC70) before the beam was delivered to the diagnostics. For temporal characterization of the pulses we used second-harmonic generation frequency-resolved optical gating (SHG FROG) with a 5-μm-thick beta barium borate (BBO) type 1 crystal. The spectra were recorded by a charge coupled device (CCD) spectrometer (AvaSpec-ULS2048XL), and the CEP stability measured using a home-built $f$-$2f$ interferometer.

Figure 1(b) shows the spectrum recorded at the fiber output when it was filled with 3.9 bar of argon and pumped by 9 fs pulses with 1 μJ pulse energy at 800 kHz repetition rate. The sharp drop in spectral power around 450 nm is due to the reflection cut-off of the chirped mirrors.

At the selected pressure the zero dispersion wavelength (ZDW) lies at 500 nm, so that the fiber provides anomalous (negative) dispersion at the pump frequency. As a result, the pump pulses undergo soliton self-compression as they propagate along the fiber with soliton order $N = \sqrt{\gamma P_0 \tau_0^2 / |\beta_2|} > 1$, where $\gamma$ is the nonlinear fiber parameter, $P_0$ is the soliton peak power, $\tau_0$ its duration and $\beta_2$ is the group velocity dispersion, which can be estimated via the capillary approximation [15]. The compression ratio and the compression length (i.e., the position along the fiber where the pulses reach their temporal focus) can be fine-tuned by varying $N$ (e.g., by varying the input energy or gas pressure) or the chirp of the pulses launched into the fiber. For the chosen parameters the soliton order is $N \sim 1.8$. This low soliton order is sufficient to self-compress the OPCPA pump pulses below two cycles while at the same time making it possible to minimize the residual uncompressed pedestal, which is higher for higher soliton orders.

Figures 2(a&b) show the measured and retrieved SHG FROG traces of the compressed pulses. The full-width at half-maximum (FWHM) pulse duration is 4.4 fs (Fig. 2(c)), with 60% of the power in the main peak. The uncompressed pedestal is mainly caused by residual higher-order dispersion in the chirped mirrors, imperfection in the initial OPCPA pulse and higher order mode contributions which will be discussed in the next paragraph. This cannot be easily compensated for and becomes particularly important as the coherent spectrum approaches bandwidths supporting pulses with durations of a few fs. The independently measured reference spectrum suggests a transform-limited pulse duration of 4.1 fs.

The retrieved spectrum (Fig. 2(d)) is periodically modulated across almost its entire bandwidth, with a fringe spacing of 12 THz. This arises from interference between the $LP_{01}$-like fundamental mode and the $LP_{11}$-like two-lobed mode. The two modes propagate through the fiber at slightly different group velocities. As a result, at the fiber output, the two modes have acquired a time delay of 90 fs, which corresponds to the fringe spacing observed in the retrieved spectrum. The interference fringes are not visible in the directly measured spectrum in Fig. 1 (b). This is because the spectrometer is usually aligned to the center of the collimated beam where the two-lobed $LP_{11}$-like mode does not carry significant power.

The presence of the higher-order mode contributes to the residual pedestal. This becomes even more evident when looking at the reconstructed spectrograms shown in Fig. 2(e&f), which were obtained at input energies of 0.7 and 1 μJ, using a 10 fs (FWHM) Gaussian pulse as the gate function.

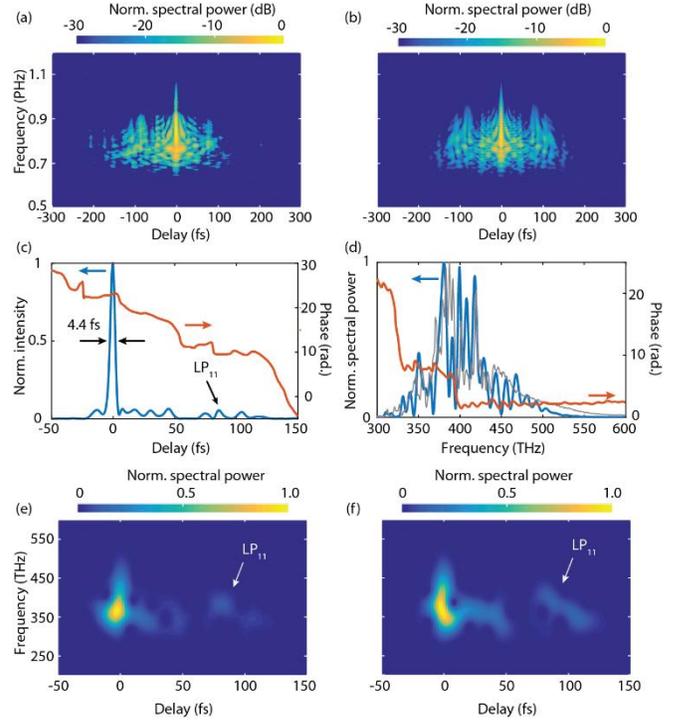

Fig. 2. (a) Measured SHG FROG trace of the self-compressed pulse at 1 μJ input energy with the spectrum shown in Fig. 1(b). (b) Retrieval of (a). (c) Retrieved temporal intensity (blue) and phase profile (orange). (d) Corresponding retrieved spectral intensity (blue) and phase (orange). The gray curve shows the reference spectrum from Fig. 1(b). (e),(f) Spectrogram of the self-compressed pulses measured at input energies of (e) 0.7 μJ and (f) 1 μJ (corresponding to the pulse shown in (c)) . Both spectrograms show a contribution of the $LP_{11}$ mode between 70-110 fs.

For the lower launched energy (Fig. 2 (e)), the self-compressed fundamental mode exits the fiber with a duration of 5.1 fs. As the pulse energy is increased (Fig. 2 (f)), the $LP_{01}$–like mode shortens further to 4.4 fs. In both cases the $LP_{11}$–like mode appears around 90 fs. The fraction of the pulse energy (between 70 fs and 110 fs) in the $LP_{11}$–like mode is 10% in Fig. 2(e) and 12.5% in Fig. 2(f).

Coupling to higher order fiber modes is caused by a spatial variation in the pulse spectrum across the beam. We discovered, by scanning the spectrometer fiber tip across the transverse position of the beam, that the OPCPA output beam has a strong spatial chirp. This effect is an intrinsic feature of the non-collinear geometry of optical parametric amplifiers [16,17] and should be taken into account in experiments where an OPCPA output is launched into a hollow-core waveguide.

We measured the CEP stability of the fiber output with a home-built $f$-$2f$ interferometer. Since the spectra were already more than one octave wide, it was sufficient to use a 500-μm-thick BBO crystal and a Glan-Thompson polarizer to combine the polarization of the fundamental signal and its second harmonic. For the measurement, supercontinuum light at 435 nm was superimposed on frequency-doubled 870 nm supercontinuum light. The resulting interference spectrum after the polarizer was measured with a CCD spectrometer (AvaSpec-ULS2048XL). A lens was used to focus the light into the spectrometer in order to take account of potential higher order mode contributions to the CEP stability. We triggered the spectrometer to the output of the OPCPA and used the shortest available integration time of 2 µs. In order to record single shot spectra, the repetition rate of the OPCPA was set to 400 kHz.

To study whether soliton self-compression preserves the CEP stability well, we compared the $f$-$2f$ measurements taken after the fiber output (Figs. 3(a-d), left column) with those at the OPCPA output, taken directly in front of the fiber after broadening the pulse spectrum via standard self-phase modulation (SPM) in a 2-mm-thick sapphire plate (Figs. 3(e-h), right column). Figs. 3(a&e) show the spectral fringes monitored over 2.5 minutes with 15 ms acquisition time. The resulting CEP variation is evaluated using Fourier transform spectral interferometry [18] yielding a root-mean-square deviation of 420 and 300 mrad (Fig. 3(b&f)). We also measured the CEP variation on a relatively short time scale of 2.5 s as shown in Figs. 3(c&g), and plotted the corresponding noise spectral densities (NSD) in Figs. 3(d&h). The upper noise frequency is limited to 240 Hz, because the fastest acquisition time of the detector is 2.1 ms, corresponding to a sampling rate of 480 Hz.

The OPCPA output exhibits lower RMS deviations from CEP stability compared to self-compressed soliton pulses. The difference arises, however, mainly from low frequency noise between 0-1.5 Hz, which we believe is caused by pointing fluctuations of the beam coupled into the fiber. By numerically cutting out frequencies below 1.5 Hz in the experimental data, the CEP noise spectral density (NSD) at the fiber output drops to the level of OPCPA output (−8 dB). We obtain RMS deviation values of 330 mrad for 15 ms acquisition time and 310 mrad for 2.1 ms, values that are comparable with the residual CEP noise of OPCPA in Fig. 3(f&g). Applying the same filter to OPCPA pulses left their RMS CEP deviations almost unchanged. A few sharp noise features in the frequency range 10 to 200 Hz are observed at both the fiber and OPCPA outputs (Fig. 3(d&h)). It is however difficult to associate these with particular noise sources because frequencies beyond 240 Hz can be aliased to the measured low frequency region.

The dependence on the input energy of the CEP of pulses undergoing spectral broadening in HC fibers was investigated in [19,20]. These works report SPM based spectral broadening in gas-filled capillaries in the normal dispersion regime and CEP characterization using $f$-$2f$ interferometry. The coupling coefficient between the measured CEP and the input energy was determined to be between 70 and 128 mrad per 1% rms energy change [19,20].

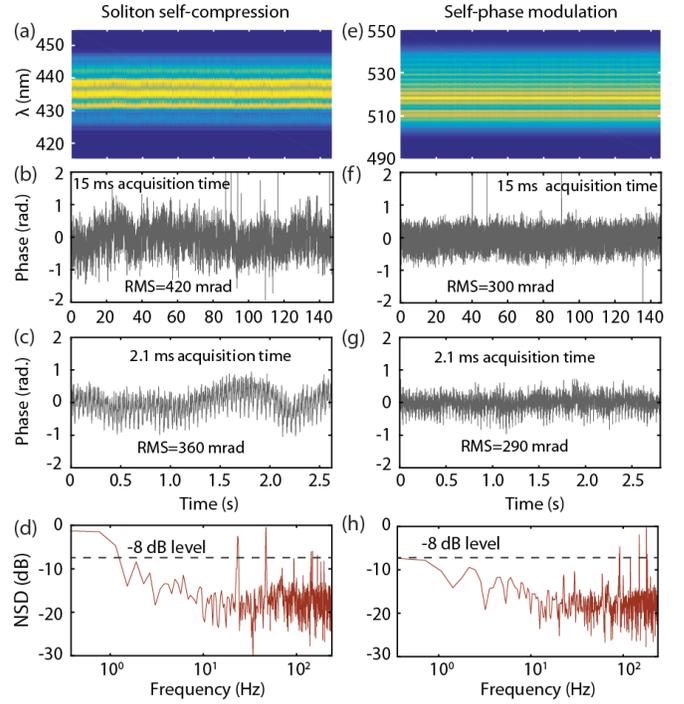

Fig. 3. CEP variation measurements of the supercontinuum obtained via soliton self-compression in the HC-PCF (left column) and via self-phase modulation in sapphire in front of the fiber (right column). (a, e) Single-shot spectral fringes measured using an $f$-$2f$ interferometer monitored over 2.5 minutes with 15 ms acquisition time and (b, f) corresponding phase drift. (c, g) Phase drift evaluated from spectral fringes monitored over 2.5 seconds with 2.1 ms acquisition time and (d, h) corresponding phase noise spectral density (NSD).

We further investigated the effect of energy fluctuations by numerical modelling, simulating pulse propagation and self-compression using the unidirectional full field equation [21]. In the simulations we used the $\chi^{(3)}$ value given in [22] and included photoionization via the Perelomov, Popov, Terent'ev model as described in [23]. We simulated the pulse evolution of an ensemble of 100 randomly distributed input energies with an RMS variation of 0.8%. These fast energy fluctuations were measured at the output of the OPCPA by collecting 160 consecutive shots with a fast photodiode and an oscilloscope. In the simulations, after propagation through the fiber, the phase noise increased by 100 mrad, in good agreement with previous results [21, 22].

It was shown in [24] that asymmetric broadening in supercontinuum generation should be avoided to preserve intrapulse coherence and to minimize CEP stability degradation. Although for the case of soliton self-compression strong self-steepening mostly extends the blue end of the spectrum, the evaluated CEP variation is similar to that of symmetric SPM supercontinuum in the sapphire plate. Accordingly, comparable CEP fluctuations indicate that soliton self-compression does not compromise intrapulse coherence but preserves the CEP just as is the case for symmetric spectral broadening via SPM. The residual slow CEP variations arise mainly from beam-pointing instabilities, which can be stabilized in future experiments.

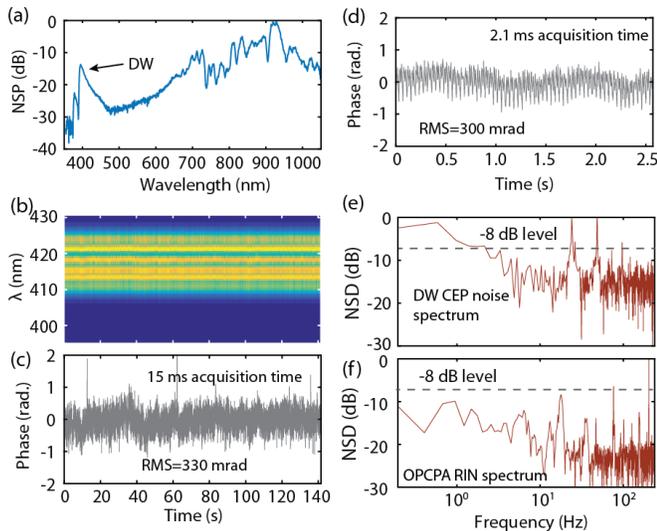

Fig. 4. (a) Experimental spectrum generated using 0.7 µJ input pulse energy at 3.6 bar of krypton with the dispersive wave around 400 nm. (b) Spectral fringes at the spectral position of the DW and (c) corresponding phase drift monitored over 2.5 minutes with 15 ms acquisition time. (d) Phase drift evaluated from spectral fringes which were monitored over 2.5 seconds with 2.1 ms acquisition time and (e) corresponding phase NSD. (f) Relative intensity noise spectral density of OPCPA output measured with the same spectrometer as for CEP variation measurements.

By changing the gas filling the fiber to 3.6 bar of krypton we were able to phase-match solitons to DWs at ~400 nm with a launched pulse energy of 0.7 µJ. The corresponding spectrum at the fiber output is shown in Fig. 4(a). DW emission has been widely exploited in both solid and hollow-core fibers for efficient and tunable frequency up- and down-conversion [25]. In HC fibers tunability from the visible down to the vacuum UV has been demonstrated at pump laser repetition rates from 1 kHz to 10 MHz [10,26–28], with conversion efficiencies up to 15% [26].

In order to characterize the DW CEP stability, we mixed the second harmonic of the pump pulse with the DW, obtaining in this way spectral fringes at ~415 nm (Fig. 4(b)). The interference pattern was monitored over 2.5 minutes, yielding 330 mrad RMS deviation of the CEP (Fig. 4(c)). A phase drift of 300 mrad RMS was also measured at the highest available sampling rate of 480 Hz. The results are shown in Fig. 4(d). The corresponding NSD in Fig. 4(e) reveals a significant low frequency component between 0 and 3 Hz with a strength greater than –8 dB. High-pass filtering with a cut-off frequency of 3 Hz reduced the RMS deviation of the CEP to 270 mrad.

Lastly, we measured the relative intensity noise (RIN) spectrum of the OPCPA with the same detector as used in the CEP measurements. The results are plotted in Fig. 4(f) for comparison. The peaks in the frequency range between 10 and 200 Hz are also observed, pointing out to the correlation between RIN and CEP NSD.

In conclusion, CEP stable 1.7 cycle pulses at 800 nm can be generated by soliton self-compression in a HC-PCF with 1 µJ of energy per pulse at 800 kHz repetition rate, providing 0.8 W of average power. Characterization of the CEP stability of the generated supercontinuum and of a DW emitted in the UV shows that the phase drift does not exceed 360 mrad, recorded in single-shot measurements over 2.5 seconds with 2.1 ms acquisition time.

The CEP variations can be reduced by improving the beam pointing stability at the input tip of the fiber and the energy stability of the OPCPA. The source is suitable for applications in strong-field physics, phase-sensitive pump-probe experiments in the UV and potentially broad UV frequency comb generation by further scaling the repetition rate of the pump laser.

**Funding.** Max Planck Society; DFG INST 90/977-1 FUGG; SFB 953; ERC grant Near Field Atto.